# Impact of Digital Time Delay on the Stable Grid Hosting Capacity of Large-scale Centralized Photovoltaic Plant

Jinhong Liu, Lin Zhou, and Marta Molinas, *Member, IEEE*

*Abstract*-- In view of the trend towards extensive application of digital controllers in the PV inverter of large-scale centralized photovoltaic (LSCPV) plant and the increasing number of grid-connected LSCPV plants, this paper investigates in detail the influence of the digital time delay of the inverter digital controller on the stable grid-hosting capacity of LSCPV plant. The studies are based on the Norton equivalent model of the grid-connected LSCPV system considering the digital time delay when modelling the digital control system of the PV inverter. Taking into account the actual situation in LSCPV plant, the stable grid-hosting capacity of LSCPV is discussed for the cases in which the PV inverters in the LSCPV plant have the same and different digital time delay values by using the root-locus method. Simulation results of the digitally controlled grid-connected LSCPV system model validates the theoretical analysis.

*Index Terms*-- Grid-connected LSCPV system, digital time delay, stable capacity, different digital time delay value.

## I. INTRODUCTION

With the advance of photovoltaic technology and continuous increase of power demand, LSCPV plants are widely employed due to their unique advantages over small-scale PV applications [1]. However, some new stability problems have emerged along with the increasing popularity of digital control system in the PV inverter of LSCPV plant and the rising of the grid penetration of LSCPV plant [2], [3].

In general, LSCPV plants are usually in remote locations, long distance transmission lines and large-capacity transformer are needed for the power transmission between LSCPV plant and load center [3]. Due to the presence of a grid impedance in the power transmission network and the paralleled-connected multi-inverter structure of LSCPV plant, the PV inverter is coupled to the grid impedance through the PCC (Point of Common Coupling) when the LSCPV is connected into the power system [4]. In the investigation of the stability of LSCPV system, Zhou *et al*. [5] analyzed the influence of the grid-impedance on the stability range of grid-connected PV inverter according to their numbers. For the multi-inverter system in LSCPV plant, Lazzarin *et al*. [6] discussed the current circulation between the inverters in a multi-inverter system and the stability of the grid-connected multi-inverter system was also analyzed in [7] and [8]. However, these studies have not considered the digital time delay of the inverter in the modeling and analysis. However, a stable system that did not consider the digital time delay in the modelling, may become unstable when the digital time delay is considered [9]. Therefore, facing the wide application of digital controller in the inverters will make inevitable the effects of digital time delay on the stability of grid-connected inverters [10]. Although several authors have addressed the problem of grid-connected multi-inverter systems, their results may not be suitable for the digitally controlled grid-connected multi-inverter systems since the digital time delay has not been taken into account in the process of modeling and analysis.

To analyze the stability of digitally controlled LSCPV system more accurately, Agorreta *et al*. [4] modeled a PV plant as a multivariable system considering the digital time delay of PV inverter and described the coupling effect existing in PV plants. The influence of the grid-connected PV plant on the grid and the stability of grid-connected PV plant due to the grid impedance are investigated in [11] and [12], respectively. Meanwhile, Luo *et al*. [13] analyzed the harmonic output characteristics of grid-connected large-scale PV plant through its harmonic transmission model that take into account the digital time delay of PV inverter. Liu *et al* [14] found that the digital time delay has a negative effect on the stability range of grid-connected PV inverter depending on the number of connected inverters based on a Norton equivalent model of digitally controlled grid-connected LSCPV system. However, this work did not provide a comprehensive analysis of the effects of digital time delay on the system. Moreover, as the main structure of LSCPV plant, the multi-inverter system has drawn more attention. On the premise that the digital time delay of the control system in PV inverter is considered, [15]-[19] studied the resonance interaction of a grid-connected multi-inverter system. In addition to the studies in the above, a detailed analysis of the stability and bandwidth limitations that occurs in digitally controlled grid-connected parallel-connected inverters is also provide in [20] and the circulating

This work was supported by the National Nature Science Foundations of China (Grant no. 51477021) and the National "111" Project of China (Grant no. B08036).

Jinhong Liu and Lin Zhou are with the Electrical Engineering Department, Chongqing University, Chongqing, 400044 China (e-mail: jinhongliu.felix@gmail.com; zhoulin@cqu.edu.cn).

Jinhong Liu and Marta Molinas are now with the Department of Engineering Cybernetics, Norwegian University of Science and Technology, Trondheim, 7491 Norway (e-mail: jinhongliu.felix@gmail.com; marta.molinas@ntnu.no).

current in the multi-inverter system is studied in [21] and [22]. Zhang *et al.* [23] have shown that increase of digital time delay of the inverter has an influence on power-sharing performance in grid-connected parallel operation of inverters. Yan *et al.* [24] discovered that the digital time delay of inverter may influence the effect of grid voltage feedback and thus reduce the effectiveness of harmonic compensation. Although many studies have been carried on the analysis of the digitally controlled grid-connected multi-inverter system, only few addressed the effects of digital time delay of inverter on grid-connected multi-inverter system. In particular, the influence of the digital time delay of inverters on the stable grid-hosting capacity of grid-connected multi-inverter system which is important to the stability of grid-connected LSCPV plant, has not been addressed.

This paper focuses on analyzing the effects of digital time delay on the stable grid-hosting capacity of grid-connected LSCPV system from the aspect of the stability range of grid-connected PV inverters depending on the number on inverters connected. In Section II, a Norton equivalent model of the grid-connected LSCPV system is built based on a typical topology of grid-connected LSCPV system considering the digital time delay of the PV inverter. The effects of the digital time delay on the stable grid-hosting capacity of LSCPV plant are studied for two situations of LSCPV plant that are described in Section III and Section IV, respectively. The simulation results are presented in Section V to validate the analysis. Finally, Conclusions are drawn in Section VI.

## II. GRID-CONNECTED LSCPV SYSTEM

### A. System Description

Fig. 1(a) shows a typical main topology of grid-connected LSCPV system, which mainly consists of the LSCPV plant and the transmission network. $T_N$ is the $N$th split-winding transformers in LSCPV plant and two PV inverters constitute an inverter module. $T_s$ represents the large-capacity step-up transformer.

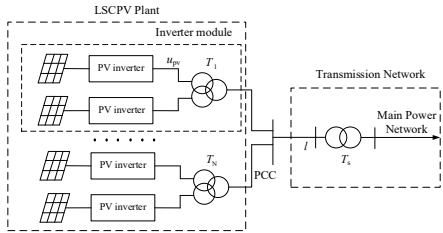

Fig. 1 Main topology of the grid-connected LSCPV system.

Generally, a cascade control strategy, which contains an inner loop and an outer loop, is used in grid-connected PV inverter [4]. The inner loop and the outer loop control the grid-side current and the dc-side voltage, respectively. Because the dynamics of the outer loop are much slower than those of the inner loop, the inner loop and the outer loop are decoupled and can be analyzed independently [5]. This manuscript mainly investigates the control of the grid-side current and thus the outer loop is not considered here. Consequently, the dc-bus of PV inverter is usually assumed as a constant voltage source.

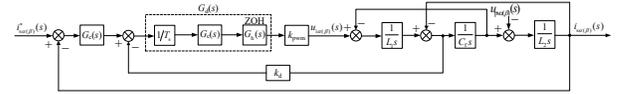

Fig. 2 Control model of the digitally controlled inverter with capacitor-current feedback active damping.

### B. Modeling of the System

To acquire an independent control of two phase current without any decoupling module, a current control loop in stationary $\alpha\beta$-frame is adopted and the control model of the digitally controlled inverter is given in Fig. 2. $G_c(s)$ is the current controller and a proportional-resonant controller is adopted.

$$G_c(s) = k_p + \frac{2k_r\omega_i s}{s^2 + 2\omega_i s + \omega_0^2} \quad (1)$$

where, $\omega_0$ is the fundamental angular frequency and $\omega_i$ represents the resonant cutoff angular frequency.

$k_{pwm}$ is the transfer function of the pulse width modulation (PWM) inverter, and $k_{pwm}=V_{dc}/2$ for the sinusoidal PWM (SPWM). $k_d$ is the damping coefficient of the active damping loop. For the digitally controlled system, the digital time delay is inevitable. $1/T_s$ represents the sampler and $T_s$ is the system sampling period. $G_d(s)$ is the computation delay and can be expressed as [11]

$$G_c(s) = e^{-s\lambda T_s} \quad (2)$$

where $\lambda$ ($0 < \lambda \leq 1$) represents the delay coefficient and $\lambda=1$ for the synchronous sampling case is adopted in the next analysis due to the wide application of the synchronous sampling in grid-connected inverters.

$G_h(s)$ represents the PWM delay, which can be modeled as zero-order holder (ZOH) and is given by [9]

$$G_d(s) = \frac{1-e^{-sT_s}}{s} \approx T_s e^{-0.5sT_s} \quad (3)$$

Then, the digital time delay in the control loop that adopted the synchronous sampling can be derived as

$$G_d(s) = \frac{1}{T_s}G_c(s)G_h(s) = e^{-1.5T_s} \quad (4)$$

It can be noted from (4) that, the digital time delay value in the control loop is $1.5T_s$. To obtain a rational transfer function, a second-order *padé* approximation is used to describe the exponential delay form shown in (4), which can be given by

$$G_{td}(s) \approx \frac{(1.5sT_s)^2 - 9T_s s + 12}{(1.5sT_s)^2 + 9T_s s + 12} \quad (5)$$

The split-winding transformer in $m$th inverter module can be equivalent two equal inductance [14], which is denoted as $L_{Tm1}=L_{Tm2}$. Considering the stability of the main power network, and the characteristics of large-capacity step-up transformer and transmission line, the transmission network is equivalent to a constant voltage ($u_g$) source in series with an impedance ($Z_g$). According to Fig. 2, an A-phase equivalent Norton model of grid-connected LSCPV system with $N$ inverter module can be used to analyze the stability of grid-connected LSCPV system because the three-phase model of grid-connected LSCPV system is interphase decoupled, which is shown in Fig. 3 [8], where $i_{pvmn}$ and $Y_{pvmn}$ is given by

$$i_{pvmn}(s) = \frac{G_{mn}(s)i_{smn}^*(s)}{Y_{eqmn}(s)L_{Tmn}s+1} \quad (6)$$

$$Y_{pvmn}(s) = \frac{Y_{eqmn}(s)}{Y_{eqmn}(s)L_{Tmn}s+1} \quad (7)$$

where $G_{smn}(s) = k_{pwm}G_{dmn}(s)$, $\omega_{rmn} = \sqrt{1/(L_{2mn}C_{fmn})}$, $\omega_{resmn} = \sqrt{(L_{1mn}+L_{2mn})/(L_{1mn}L_{2mn}C_{fmn})}$,

$$G_{mn}(s) = \frac{G_{smn}(s)\omega_{rmn}^2 G_{cmn}(s)}{L_{1mn}s^3 + k_{dmn}G_{smn}(s)s^2 + L_{1mn}\omega_{resmn}^2 s + G_{smn}(s)G_{cmn}(s)\omega_{rmn}^2} \quad (8)$$

$$Y_{eqmn}(s) = \frac{(L_{1mn}/L_{2mn})s^2 + (k_{dmn}G_{smn}(s)/L_{2mn})s + \omega_{rmn}^2}{L_{1mn}s^3 + k_{dmn}G_{smn}(s)s^2 + L_{1mn}\omega_{resmn}^2 s + G_{smn}(s)G_{cmn}(s)\omega_{rmn}^2} \quad (9)$$

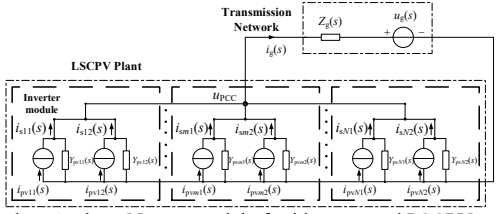

Fig. 3 Equivalent A-phase Norton model of grid-connected LSCPV system.

Based on Fig. 3, the transfer function of grid-side current of $n$th inverter in $m$th inverter module can be derived as

$$i_{smn}(s) = (1 - \frac{Y_{pvmn}(s)}{\sum_{m=1}^{N}\sum_{n=1}^{2}Y_{pvmn}(s) + Y_g(s)})i_{pvmn} - \frac{Y_{pvmn}(s)Y_g(s)}{\sum_{m=1}^{N}\sum_{n=1}^{2}Y_{pvmn}(s) + Y_g(s)}u_g(s) \\ - \sum_{t=1}^{N}\sum_{h=1,h\neq n}^{2}\frac{Y_{pvmn}(s)}{\sum_{m=1}^{N}\sum_{n=1}^{2}Y_{pvmn}(s) + Y_g(s)}i_{pvth}(s) \quad (10)$$

### III. STABLE GRID-HOSTING CAPACITY OF THE LSCPV PLANT WITH SAME DIGITAL TIME DELAY VALUE

For a LSCPV plant, the stable grid-connected capacity of LSCPV plant concerns the stable operation of grid-connected LSCPV system and the design of grid-connected LSCPV plant [12]-[14]. Thus, it is necessary to analyze and study the effects of digital time delay on the stable grid-hosting capacity of LSCPV plant to ensure the realization of stable grid-connection with allowable grid-connected capacity under a certain grid situation. Meanwhile, because the stability range of grid-connected inverter depending on the number of connected inverters is directly related to the stable grid-hosting capacity of the LSCPV plant, the analysis of the effects of digital time delay on the stable grid-hosting capacity of the LSCPV plant can be translated into the analysis of the effects of the digital time delay on the stability range of PV inverter numbers in grid-connected LSCPV plant.

In the analysis of the grid-connected LSCPV system, the PV inverter in LSCPV plant is usually assumed to be stable before connecting to the grid-connected parallel inverter system of LSCPV plant. According to (6) and the parameters of 500 kW PV inverter that listed in Table I, the root locus of grid-side current with digital time delay changes is shown in Fig. 4. As seen, the increase of digital time delay will undermine the stability of PV inverter and the stability range of the digital time delay value can be derived as (11), by calculating the intersections of the root locus of (6) and imaginary axis.

TABLE I
MAIN PARAMETERS OF THE 500 KW INVERTERS IN LSCPV PLANT

| Parameters | Values | Parameters | Values |
| --- | --- | --- | --- |
| $k_d$ | 0.001 | $L_1$ | 90 μH |
| $k_r$ | 1 | $L_2$ | 18 μH |
| $k_d$ | 0.0017 | $C_f$ | 182 μF |
| $\omega_0$ | 100π rad/s | $\omega_i$ | π rad/s |
| $V_{dc}$ | 553 V | $u_{pv}$(rms) | 156 V |
| $f_{sw}$(Switching frequency) | 10 kHz | $f_s$ | 20 kHz |

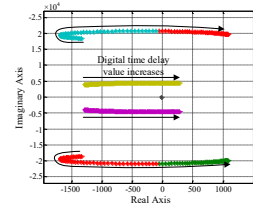

Fig. 4 Root locus of PV inverter as digital time delay value increases.

$$0 \text{ μs} < T_d < 85.5 \text{ μs} \quad (11)$$

In most LSCPV plants, the PV inverters are installed by the same firm and are of the same type [4]. Hence, the parameters of the PV inverters in LSCPV plant can be assumed the same and then (10) can be simplified as

$$i_{smn}(s) = (1 - \frac{Y_{pv}(s)}{2NY_{pv}(s) + Y_g(s)})i_{pvmn} - \frac{Y_{pv}(s)Y_g(s)}{2NY_{pv}(s) + Y_g(s)}u_g(s) \\ - \sum_{t=1}^{N}\sum_{h=1,h\neq n}^{2}\frac{Y_{pv}(s)}{2NY_{pv}(s) + Y_g(s)}i_{pvth}(s) \quad (12)$$

The stability of grid-connected LSCPV system depends on the stability of grid-side current. Substituting (6) and (7) to (12), the closed-loop transfer function of grid-side current of inverter in LSCPV plant can be derived as

$$i_{smn}(s) = \left[1 - \frac{Y_{eq}(s)}{2NY_{eq}(s) + Y_g(s)(Y_{eq}(s)L_Ts+1)}\right] \cdot \frac{G(s)}{Y_{eq}(s)L_Ts+1}i_{smn}^*(s) \\ - \frac{Y_{eq}(s)Y_g(s)}{2NY_{eq}(s) + Y_g(s)(Y_{eq}(s)L_Ts+1)}u_g(s) \\ - \frac{Y_{eq}(s)}{2NY_{eq}(s) + Y_g(s)(Y_{eq}(s)L_Ts+1)} \cdot \frac{G(s)}{Y_{eq}(s)L_Ts+1}\sum_{t=1}^{N}\sum_{h=1,h\neq n}^{2}i_{sth}^*(s) \quad (13)$$

Fig. 5 shows the root locus of $i_{smn}$ when the number of grid-connected PV inverters with different digital time delay value varies. As seen, the closed-loop poles gradually approach and finally cross the imaginary axis as the number of grid-connected PV inverter increases and the poles will approach the imaginary axis again when the number of grid-connected inverters continue to increase. This indicates that the system stability firstly decreases and then increases when the number of grid-connected inverter continue to increase. Moreover, the increased digital time delay value enable poles to get closer to imaginary axis, which means the stability of system decreases with increasing digital time delay value.

According to the stability criterion, the intersections of the root locus and the imaginary axis are the stability boundary of the number of grid-connected inverter. Table II shows the stability ranges of the number of grid-connected PV inverters in LSCPV plant when digital time delay value of PV inverters is different, which are derived from the characteristic equation of (13). It can be found that, the stability range of the grid-connected PV inverter number decreases significantly with the introduction of the digital time delay and the stability range of grid-connected inverter number decrease with the increase of digital time delay value. It means that digital time delay will dramatically reduce the stable grid-hosting capacity of LSCPV plants and the integration capacity of LSCPV plant will decrease as the digital time delay value increases.

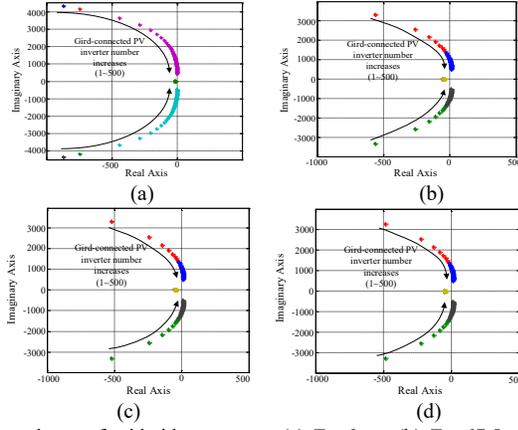

Fig. 5 Root locus of grid-side current $i_s$. (a) $T_d= 0$ μs, (b) $T_d= 67.5$ μs, (c) $T_d= 75$ μs, (d) $T_d= 82.5$ μs.

TABLE II
STABILITY RANGES OF THE GRID-CONNECTED PV INVERTER NUMBER

| Digital time delay ($T_d$) value | Stability ranges of grid-connected PV inverter number ($2N$) |
|---|---|
| 0 μs | $2N < 520$ |
|  | $2N > 566$ |
| 67.5 μs | $2N < 48$ |
|  | $2N > 613$ |
| 75 μs | $2N < 20$ |
|  | $2N > 708$ |
| 82.5 μs | $2N < 9$ |
|  | $2N > 794$ |

## IV. STABLE GRID-HOSTING CAPACITY OF THE LSCPV PLANT WITH DIFFERENT DIGITAL TIME DELAY VALUES

In practical applications, different types of PV inverter may be adopted in the extension of LSCPV plant and thus this may result in a situation of the PV inverters in LSCPV plant having different digital time delay values. Assuming the PV inverters in LSCPV plant contains $h$ ($h>1$) different values of digital time delay, and the number of the PV inverter with the same digital time delay value is denoted as $N_i$ ($i=1\sim h$). According to (10), the transfer function of the grid-side current of $j$th ($0 < j < N_i$) inverter in $h$ size inverters can be derived as

$$i_{sij}(s) = \left(1 - \frac{Y_{pvij}(s)}{\sum_{k=1}^{h}\sum_{l=1}^{N_i} Y_{pvkl}(s) + Y_g(s)}\right) i_{pvij}(s) - \frac{Y_{pvij}(s)Y_g(s)}{\sum_{k=1}^{h}\sum_{l=1}^{N_i} Y_{pvkl}(s) + Y_g(s)} u_g(s) \quad (14)$$
$$- \sum_{q=1, q\neq j}^{N_i} \frac{Y_{pvij}(s)}{\sum_{k=1}^{h}\sum_{l=1}^{N_i} Y_{pvkl}(s) + Y_g(s)} i_{pviq}(s) - \sum_{p=1, p\neq i}^{h}\sum_{q=1}^{N_p} \frac{Y_{pvij}(s)}{\sum_{k=1}^{h}\sum_{l=1}^{N_i} Y_{pvkl}(s) + Y_g(s)} i_{pvpq}(s)$$

Considering the transfer function of equivalent current source ($i_{pv}$) and equivalent admittance ($Y_{pv}$) are the same for the PV inverters with the same digital time delay value. Thus, the grid-side current of the PV inverter with the same digital time delay value are the same and its function can be obtained from (14) as

$$i_{si}(s) = \left(1 - \frac{N_i Y_{pvi}(s)}{\sum_{k=1}^{h} N_k Y_{pvk}(s) + Y_g(s)}\right) i_{pvi}(s) - \frac{Y_{pvi}(s)Y_g(s)}{\sum_{k=1}^{h} N_k Y_{pvk}(s) + Y_g(s)} u_g(s) \quad (15)$$
$$- \sum_{p=1, p\neq i}^{h} \frac{N_p Y_{pvi}(s)}{\sum_{k=1}^{h} N_k Y_{pvk}(s) + Y_g(s)} i_{pvp}(s)$$

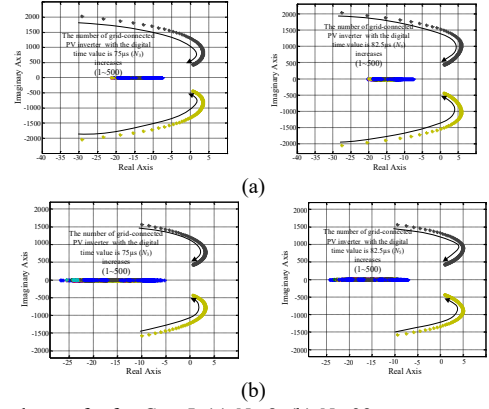

Fig. 6 Root locus of $i_s$ for Case I. (a) $N_1=8$. (b) $N_1=32$.

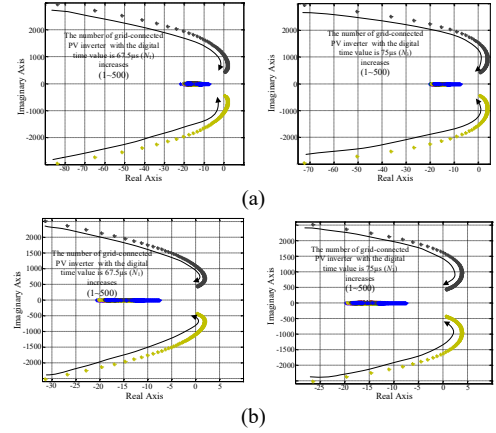

Fig. 7 Root locus of $i_s$ for Case II. (a) $N_5=2$. (b) $N_5=8$.

In order to facilitate the analysis, the numbers of the PV inverters with digital time delay values of 67.5 μs, 72 μs, 75 μs, 79.5 μs and 82.5 μs are denoted as $N_1$, $N_2$, $N_3$, $N_4$ and $N_5$, respectively. Considering that the PV inverters in the original LSCPV plant have the same digital time delay value, which means there is only one size of digital time delay value in the PV inverters of LSVPV plant, two cases are divided as follows:

*Case* I: When the digital time delay value of PV inverters is 67.5 μs in the original LSCPV plant and the number of such PV inverter is 8 and 32, respectively, the inverters with a larger digital time delay value are added to the LSCPV plant.

*Case* II: When the digital time delay value of PV inverter is 82.5 μs in the original LSCPV plant and the number of such PV inverters is 2 and 8, respectively, the PV inverters with a smaller digital time delay value are added to the LSCPV plant.

According to (15), the root locus of grid-side current for *Case* I and *Case* II as the added PV inverter number increases,

TABLE III
STABILITY RANGES OF THE GRID-CONNECTED PV INVERTER NUMBER FOR
CASE I AND CASE II

| Cases | Stability ranges of the grid-connected PV inverters number | | |
|---|---|---|---|
| | 67.5 μs | 75 μs | 82.5 μs |
| Case I | $N_1$=8(original PV inverter number) | $N_3$< 18, $N_3$> 699 | $N_5$< 10, $N_5$> 784 |
| | $N_1$=32(original PV inverter number) | $N_3$< 8, $N_3$> 672 | $N_5$< 6, $N_5$> 755 |
| Case II | $N_1$< 42, $N_1$> 612 | $N_3$< 16, $N_3$> 706 | $N_5$=2 (original PV inverter number) |
| | $N_1$< 22, $N_1$> 608 | $N_3$< 3, $N_3$> 701 | $N_5$=8(original PV inverter number) |

are shown in Fig. 6 and Fig. 7, respectively. As seen, there also exists a pair of poles that varies with the grid-connected PV inverter number. Based on the stability criterion, the intersections of the root locus and the imaginary axis in Fig. 6 and Fig. 7 are the marginal number of the PV inverter that can be connected to the grid before stability is lost, and this marginal number is decreased as the digital time delay value of added inverter increases. Table III shows the stability ranges of grid-connected PV inverter in Case I and Case II.

As seen in Table III, the stability ranges of the grid-connected PV inverter number will decrease with the increase of the digital time delay value of the added PV inverter. By comparing the stability ranges of grid-connected PV inverter listed in Table II and Table III, it can be found that addition of PV inverters with a larger digital time delay value will reduce the stability range of the grid-connected PV inverter number, and the stability range of the grid-connected PV inverter number will be extended as the PV inverters that have a smaller digital time delay are added. After considering the number of the PV inverter in original LSCPV plant, the stability ranges of the total grid-connected PV inverter number is increased in Case I and decreased in Case II as the number of PV inverter in original LSCPV plant increases.

In addition to the situation of the PV inverters in the original LSCPV plant having the same digital time delay value, the situation of the PV inverters in the original LSCPV plant having different digital time delay values is also discussed based on three different cases given below:

*Case* III: The PV inverters in the original LSCPV plant have two different values of digital time delay: 67.5 μs and 75μs. When the number of the PV inverters that have the same digital time delay value in the original LSCPV plant are $N_1$=2, $N_3$=6 and $N_1$=6, $N_3$=2, respectively, PV inverters that have a larger digital time delay value are added.

*Case* IV: The PV inverters in the original LSCPV plant have two different values of digital time delay: 67.5 μs and 82.5μs. When the number of PV inverters that have the same digital time delay value in the original LSCPV plant are $N_1$=2, $N_5$=6 and $N_1$=6, $N_5$=2, respectively, the added PV inverters have a digital time delay value that is midway between two digital time delay values of the PV inverters in the original LSCPV plant.

*Case* V: The PV inverters in the original LSCPV plant have two different values of digital time delay: 75 μs and 82.5μs. When the number of PV inverters that have the same digital

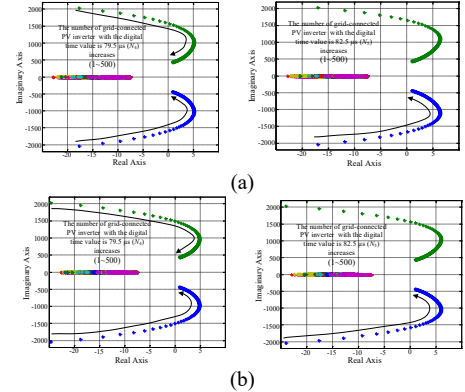
(a)

(b)
Fig. 8 Root locus of $i_s$ for Case III. (a) $N_1$=2, $N_3$=6. (b) $N_1$=6, $N_3$=2.

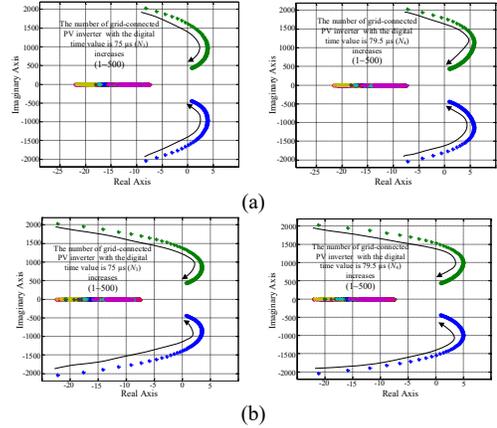
(a)

(b)
Fig. 9 Root locus of $i_s$ for Case IV. (a) $N_1$=2, $N_5$=6. (b) $N_1$=6, $N_5$=2.

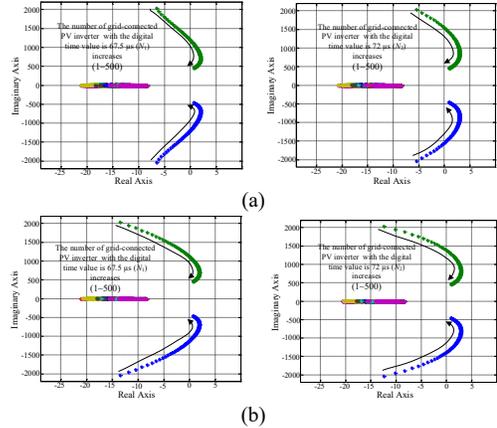
(a)

(b)
Fig. 10 Root locus of $i_s$ for Case V. (a) $N_3$=2, $N_5$=6. (b) $N_3$=6, $N_5$=2.

time delay value in the original LSCPV plant are $N_3$=2, $N_5$=6 and $N_3$=6, $N_5$=2, respectively, PV inverters that have a smaller digital time delay values are added.

Fig. 8 to Fig. 10 show the root locus of grid-side current $i_s$ for the above three cases. It can be seen from Fig. 8 to Fig. 10 that, the marginal numbers of PV inverters that can be connected to the grid before stability is lost is decreased when the digital time value of added PV inverter increases. When the LSCPV plant has a larger number of PV inverters with smaller digital time delay value, the marginal number of PV inverters that can be grid-connected stably also become larger

Table IV shows the stability ranges of grid-connected PV inverter number for each case, which are obtained by calculating the intersections of the root locus of (17) and the imaginary axis in $s$ domain. The stability ranges of the PV inverter number listed in Table IV further illustrates that the stability ranges of grid-connected PV inverter number decreases with the increase of digital time delay values of added inverters. An original LSCPV plant that has more PV inverters that have a smaller digital time delay value than other PV inverters will have greater stability ranges of grid-connected PV inverter numbers.

TABLE IV
STABILITY RANGES OF THE GRID-CONNECTED PV INVERTER NUMBER FOR CASE III, CASE IV AND CASE V

| Cases | Original PV inverter in LSCPV plant | Stability ranges of the grid-connected PV inverters number | | | | |
|---|---|---|---|---|---|---|
| | | 67.5 μs | 72 μs | 75 μs | 79.5 μs | 82.5 μs |
| Case III | 67.5 μs, $N_1$=2; 75 μs, $N_3$=6 | - | - | - | $N_4$< 9, $N_4$> 751 | $N_5$< 7, $N_5$> 784 |
| | 67.5 μs, $N_1$=6; 75 μs, $N_3$=2 | - | - | - | $N_4$< 12, $N_4$> 751 | $N_5$< 9, $N_5$> 784 |
| Case IV | 67.5 μs, $N_1$=2; 82.5 μs, $N_5$=6 | - | - | $N_3$< 8, $N_3$> 700 | $N_4$< 5, $N_4$> 752 | - |
| | 67.5 μs, $N_1$=6; 82.5 μs, $N_5$=2 | - | - | $N_3$< 15, $N_3$> 699 | $N_4$< 10, $N_4$> 751 | - |
| Case V | 75 μs, $N_3$=2; 82.5 μs, $N_5$=6 | $N_1$< 25, $N_1$> 607 | $N_2$< 10, $N_2$> 665 | - | - | - |
| | 75 μs, $N_3$=6; 82.5 μs, $N_5$=2 | $N_1$< 30, $N_1$> 607 | $N_2$< 15, $N_2$> 664 | - | - | - |

Therefore, for the case when the original PV inverters contain more than one value of digital time delay, (e.g. the PV inverters in original LSCPV plant have multiple different digital time delay value), the PV inverter number that can be connected to the grid stably will decrease as the digital time delay value of added PV inverters increases. The addition of the PV inverters with a smaller digital time delay value will increase the stable grid-connected PV inverter number and that increasing tendency will become more obvious as the digital time delay value of added PV inverters decreases. Meanwhile, the stability range of the grid-connected PV inverter number in LSCPV plant becomes larger when there are more PV inverters that have a smallest digital time delay value in the original LSCPV plant. Moreover, compared with the situations where the added PV inverters have a larger and smaller digital time delay value than the original PV inverters, the addition of PV inverters with an intermediate digital time delay value relative to the digital time delay values in the original LSCPV plant PV inverters will contribute to a greater stability ranges of grid-connected PV inverters numbers in the LSCPV plant.

## V. SIMULATION RESULTS

In order to validate the theoretical analysis and the obtained stability ranges of grid-connected PV inverter numbers which are affected by the digital time delay of PV inverter, a simulation model of grid-connected LSCPV system was built in MATLAB/SIMULINK. The main parameters of PV inverter and other main parts are shown in Table I and Table V, respectively.

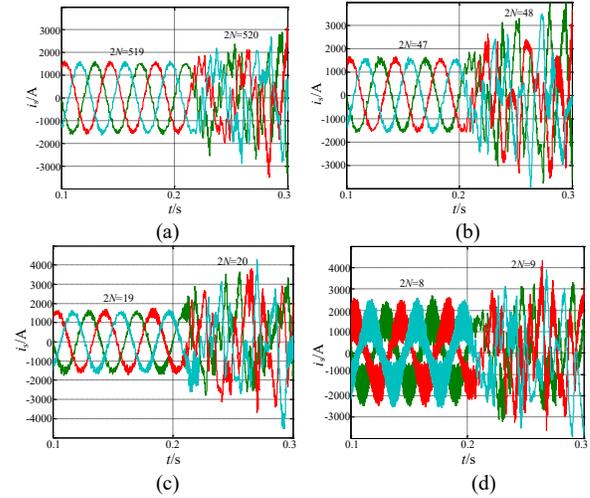

Fig. 11 Simulation waveforms of $i_s$ under different digital time delay value $T_d$. (a) $T_d$=0 μs. (b) $T_d$=67.5 μs. (c) $T_d$=75 μs. (d) $T_d$=82.5 μs.

TABLE V
MAIN PARAMETERS OF THE DOUBLE SPLIT WINDING AND TRANSMISSION NETWORK

| Component | Description | Notation | Value |
|---|---|---|---|
| Double split winding transformer | Rated voltage ratio (High voltage/Low voltage 1/Low voltage 2) | $U_h/U_{l1}/U_{l2}$ | 10±2×2.5%/0.27/0.27 kV |
| | Impedance voltage | $U_z$ | 4.5% |
| | Rated capacity | $S_s$ | 1000/500/500 kV |
| Large-capacity step-up transformer | Short circuit voltage ratio (%) | $U_s$% | 10.5 |
| | Rated voltage of high voltage side | $U_H$ | 110 kV |
| | Rated voltage of low voltage side | $U_L$ | 10 kV |
| | Rated capacity | $S$ | 6300 kVA |
| Transmission line | Unit resistance | $r$ | 0.21 Ω/km |
| | Unit reactance | $x$ | 0.34 Ω/km |
| | Line length | $l$ | 20 km |

Fig. 11 shows the grid-side current of PV inverter when increase the PV inverters in LSCPV plant have the same digital time delay values. As seen, the grid-side current of PV inverter is stable only if the number of grid-connected PV inverter within the stability ranges listed in Table II and the number of the PV inverters that can be connected to grid stably is decreased as the digital time delay value of PV inverter increases.

Fig.12 to Fig.16 show the grid-side current of PV inverter for the five previous given *Cases* when the grid-connected PV inverter number changes. It can be found that grid-side current of PV inverter will become unstable when the grid-connected PV inverter number is outside the stability ranges listed in Table III and Table IV.

Fig. 12 and Fig. 13 show the simulation waveforms of grid-side current of PV inverter in Case I and Case II, respectively. As seen, the stability ranges of grid-connected PV inverter number decrease as the addition of the PV inverters have a larger digital time delay value than the PV inverters in original LSCPV plant. On the contrary, adding PV inverters with smaller digital time delay values than the PV inverter in original LSCPV plant will increase the stability ranges of grid-connected PV inverter number.

It can be seen from Fig. 14 to Fig. 16 that, the simulation waveforms of the grid-side current of PV inverter in *Case* III, *Case* IV and *Case* V prove that stability ranges of grid-connected PV inverter number will become larger when the original LSCPV have a larger number of PV inverter with smaller digital time delay value than the other PV inverters. Meanwhile, when the PV inverter number in original LSCPV plant is the same, the situation of the addition of the PV inverters with smaller digital time delay value than the PV inverter in the original LSCPV plant has a much wider stability ranges of grid-connected PV inverter number.

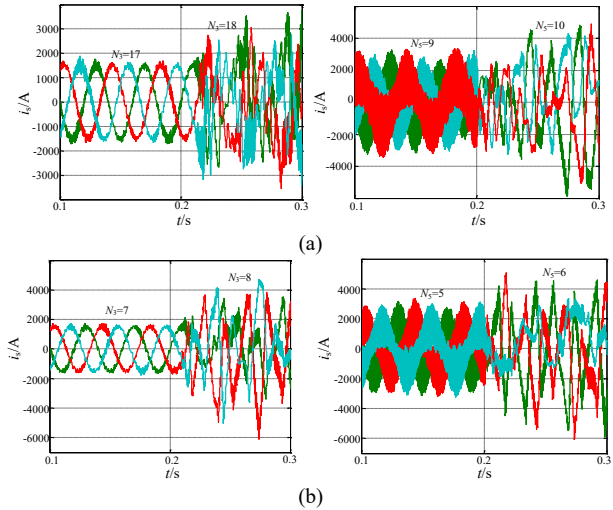

Fig. 12 Simulation waveforms of $i_s$ for *Case* I. (a) $N_1$=8. (b) $N_1$=32.

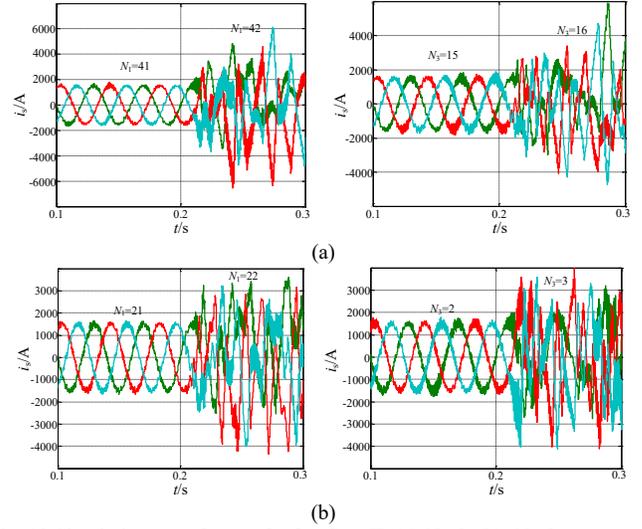

Fig. 13 Simulation waveforms of $i_s$ for *Case* II. (a) $N_5$=2. (b) $N_5$=8.

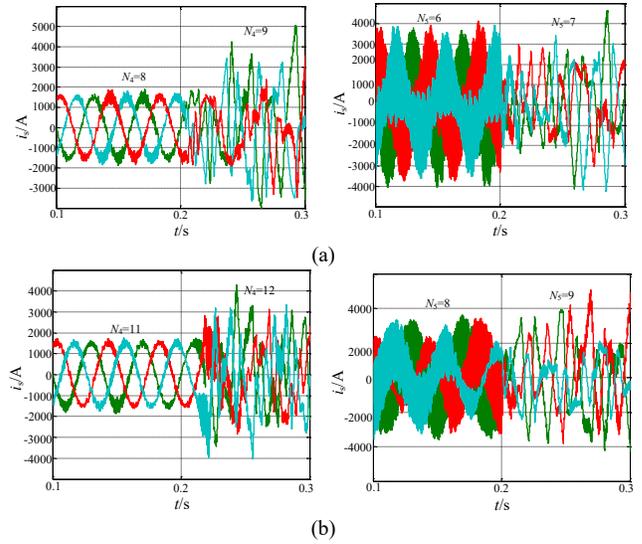

Fig. 14 Simulation waveforms of $i_s$ for *Case* III. (a) $N_1$=2, $N_3$=6. (b) $N_1$=6, $N_3$=2.

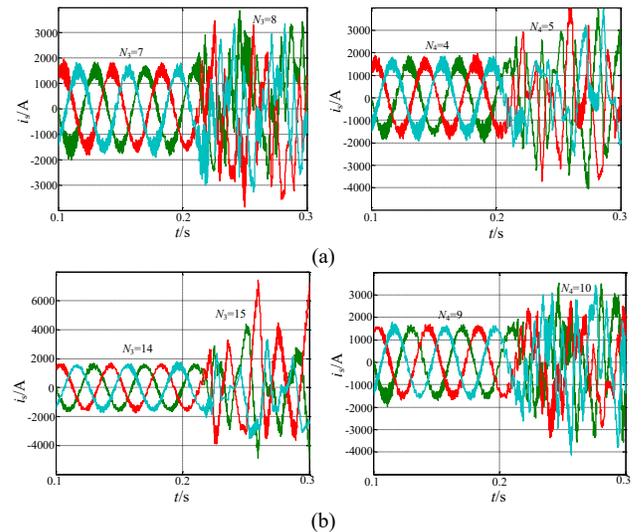

Fig. 15 Simulation waveforms of $i_s$ for *Case* IV. (a) $N_1$=2, $N_5$=6. (b) $N_1$=6, $N_5$=2.

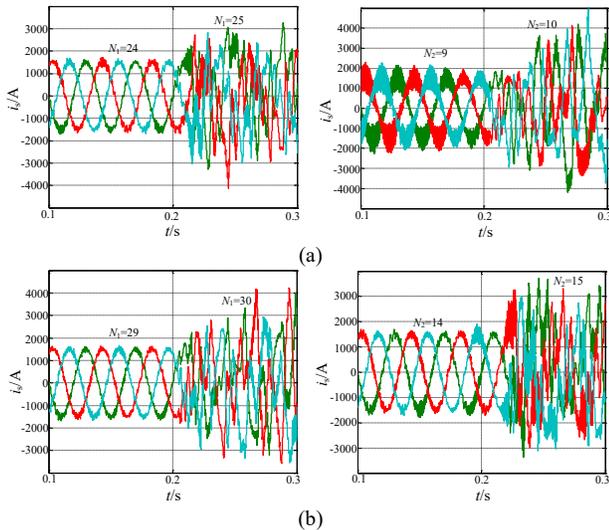

Fig. 16 Simulation waveforms of $i_s$ for *Case* V. (a) $N_3$=2, $N_5$=6. (b) $N_3$=6, $N_5$=2.

VI. CONCLUSIONS

The effects of digital time delay of PV inverters on the grid-hosting capacity of LSCPV plants has been thoroughly analyzed in this paper. The equivalent Norton model of grid-connected LSCPV system is built firstly, which is the foundation for the analysis of the connected system. Then, considering the relation between the grid-connected capacity of LSCPV plant and the numbers of grid-connected PV inverters, the influence of digital time delay of PV inverters on the stable grid-hosting capacity of LSCPV plant have been studied through the analysis of the stability ranges of grid-connected PV inverter numbers for cases where the PV inverters in the LSCPV plant have the same and different digital time delay values. The analysis results show that the digital time delay of PV inverter has a negative effect on the stable grid-hosting capacity of LSCPV plant and different proportion of PV inverters that have different values of digital time delay will also affect the stability ranges of grid-connected PV inverter numbers. The theoretical analysis results were verified through simulation results and can be used as reference for the stable grid-hosting capacity of LSCPV plants.